\documentclass[12pt,preprint]{aastex}
\usepackage{emulateapj5}
\begin{document}
                                                                                               
\newcommand{\kms}{\mbox{km~s$^{-1}$}}
\newcommand{\s}{\mbox{$''$}}
\newcommand{\mloss}{\mbox{$\dot{M}$}}
\newcommand{\my}{\mbox{$M_{\odot}$~yr$^{-1}$}}
\newcommand{\ls}{\mbox{$L_{\odot}$}}
\newcommand{\um}{\mbox{$\mu$m}}
\newcommand{\ujy}{\mbox{$\mu$Jy}}
\newcommand{\ms}{\mbox{$M_{\odot}$}}
\newcommand\mdot{$\dot{M}  $}

\title{SHOCKED AND SCORCHED: THE TAIL OF A TADPOLE IN AN INTERSTELLAR POND}
\author{R. Sahai\altaffilmark{1}, M. R. Morris\altaffilmark{2}, M.J. Claussen\altaffilmark{3}
}

\altaffiltext{1}{Jet Propulsion Laboratory, MS\,183-900, California
Institute of Technology, Pasadena, CA 91109}

\altaffiltext{2}{Department of Physics and Astronomy,
UCLA, Los Angeles, CA 90095-1547}

\altaffiltext{3}{National Radio Astronomy Observatory, 1003 Lopezville Road,
Socorro, NM 87801}

%\altaffiltext{4}{Dpto. de Astrofisica Molecular e
%Infraroja, Instituto de Estructura de la Materia-CSIC, Serrano 121, 
%28006 Madrid, Spain}
\email{raghvendra.sahai@jpl.nasa.gov}
                                                                                             
\begin{abstract}
We report multi-wavelength observations of the far-infrared source IRAS\,20324+4057, including high-resolution optical imaging with HST, 
and ground-based near-infrared, millimeter-wave and radio observations. These data show an extended, limb-brightened, tadpole-shaped nebula with a 
bright, compact, cometary nebula located inside the tadpole head. Our molecular line observations indicate that the Tadpole is predominantly molecular, with a total gas mass 
exceeding 3.7\,\ms. Our radio continuum imaging, and archival Spitzer IRAC images, show the presence of additional tadpole-shaped objects in the 
vicinity of IRAS\,20324+4057 that share a common E-W head-tail orientation: we propose that these structures are small, dense molecular cores that 
originated in the Cygnus cloud and are now being (i) photoevaporated by the ultraviolet radiation field of the Cyg OB2 No. 8 cluster located to the North-West, 
and (ii) shaped by ram pressure of a distant wind source or sources located to the West, blowing 
ablated and photoevaporated material from their heads eastwards. The ripples in the tail of the Tadpole are interpreted in terms of instabilities at 
the interface between the ambient wind and the dense medium of the former.
\end{abstract}
                                                                                             

\section{Introduction}

The presence of strong radiation fields from massive O and B stars can be detrimental to nearby star 
formation, when compared with star formation
in the absence of nearby high-mass stars.  Much has been published about this
phenomenon; the ionizing radiation from massive stars can evaporate circumstellar
disks around nearby low-mass stars (partially ionized globules, i.e.
PIGS; see, e.g., Laques \& Vidal 1979 and Garay 1987\footnote{Since the Garay review paper, the 
original PIGs have largely been recast as proplyds, and the acronym PIG has been used 
to describe much larger bright-rimmed molecular clouds, e.g., Serabyn, Gusten, \& Mundy 1993}) 
or can progressively ionize dense blobs of nearby molecular material, forming
evaporating gaseous globules or EGGs, defined explicitly by Hester et al. (1996) as ``globules 
of dense gas that are being photoevaporated more slowly than their 
lower density surroundings, and so are left behind as the gas around them is driven off."  These
terms, which describe the ionization/evaporation of gas in disks around low-mass
stars, or surrounding molecular material, are rather ill-defined otherwise.  The term
``proplyd'' (photo-evaporating protoplanetary disk) is probably the best defined, and is generally accepted as referring to
evaporating circumstellar disks around young, low-mass stars.  An extensive literature 
can be found about proplyds, especially those in the 
Orion Nebula Cluster (e.g. O'Dell, Wen, \& Hu 1993; Ricci, Robberto,
\& Soderblom 2008), and their interpretation (e.g. Storzer \& Hollenbach
1999; Johnstone, Hollenbach, \& Bally 1998).  Sizes and 
mass-loss rates of proplyds in the Orion Nebula Cluster range from 40--350 AU and 
0.7--1.5 $\times$10$^{-6}$\my~(Henney \& O'Dell 1999). The EGGs, 
first discovered in the Eagle Nebula, appear to be connected to ongoing star formation; 
near-infrared, high angular resolution observations show that $\sim$15\% of the EGGs
in the ``elephant trunks'' of the Eagle show evidence for associated young low-mass
stars, and that there is also evidence for relatively massive young stellar objects at the 
tips of the elephant trunks (McCaughrean \& Andersen 2002). With the advent of $Spitzer$ and the 
{\it Wide-field Infrared Survey Explorer (WISE)}, new imaging surveys 
have revealed the wide-spread presence of elephant-trunk structures (also referred to as ``pillars") and EGGs towards many massive 
star-forming regions  (e.g., Smith et al. 2010, Koenig et al. 2012).

We have serendipitously discovered an extended, limb-brightened, tadpole-shaped
nebula associated with the IRAS source IRAS\,20324+4057 (I\,20324) near the Cygnus OB2 No. 8
cluster.  This paper presents HST imaging, Palomar 5-m spectroscopy, Spitzer
IRAC and MIPS imaging, VLA radio continuum imaging, and (sub) millimeter-wave spectroscopy
using the ARO SMT and 12-m in order to elucidate the nature of this object and
its relationship to proplyds, PIGs, or EGGs.

I\,20324 was observed in an HST imaging survey of a list of candidate long-lived preplanetary nebulae (PPNs) selected from the
IRAS Point Source Catalog using a color-criterion ($F_{60}/\,F_{25}\,>\,1$) chosen to select objects that are dominated by relatively cool dust located 
at some distance from a stellar source, or where any hot dust is highly obscured (Sahai et al.
2007). All objects were required to have point-source 2MASS (JHK) and MSX (A band/8\um) counterparts, implying the presence of
local stellar sources that have heated the dust revealed in the HST and IRAS data. However, we found that a large fraction of
the objects resolved in our survey reveal morphologies quite different than those found for PPNs (e.g. Sahai
et al. 2007). I\,20324 is one of the most prominent of these. It was first detected as an emission-line nebula together with two other nebulous objects within a $3.1{'}\times3.1{'}$ field, in a ground-based 
H$\alpha+[NII]$ survey of candidate post-AGB objects (Pereira \& Miranda 2007: PM07). PM07 concluded that all 3 objects are 
``related young stellar objects" and that the emission-line spectrum of I\,20324 is indicative of photoionization rather than shock-excitation. 

Independently, I\,20324 has been noted recently by Wright et al. (2012: Wetal12), who included it among
a group of 10 ``proplyd-like objects" associated with the Cygnus OB2 association.  They 
considered both the proplyd and EGG hypotheses, and concluded that neither 
scenario adequately explains their observations.  They suggest that these and the 
other objects they identified are ``a unique class of  photoevaporating partially 
embedded young stellar objects".  In contrast, we find that our data favor the
EGG hypothesis. 

\section{Observations \& Results}
\subsection{HST imaging}\label{sub_hst}
\begin{figure}[!h]
\resizebox{1.0\textwidth}{!}{\includegraphics{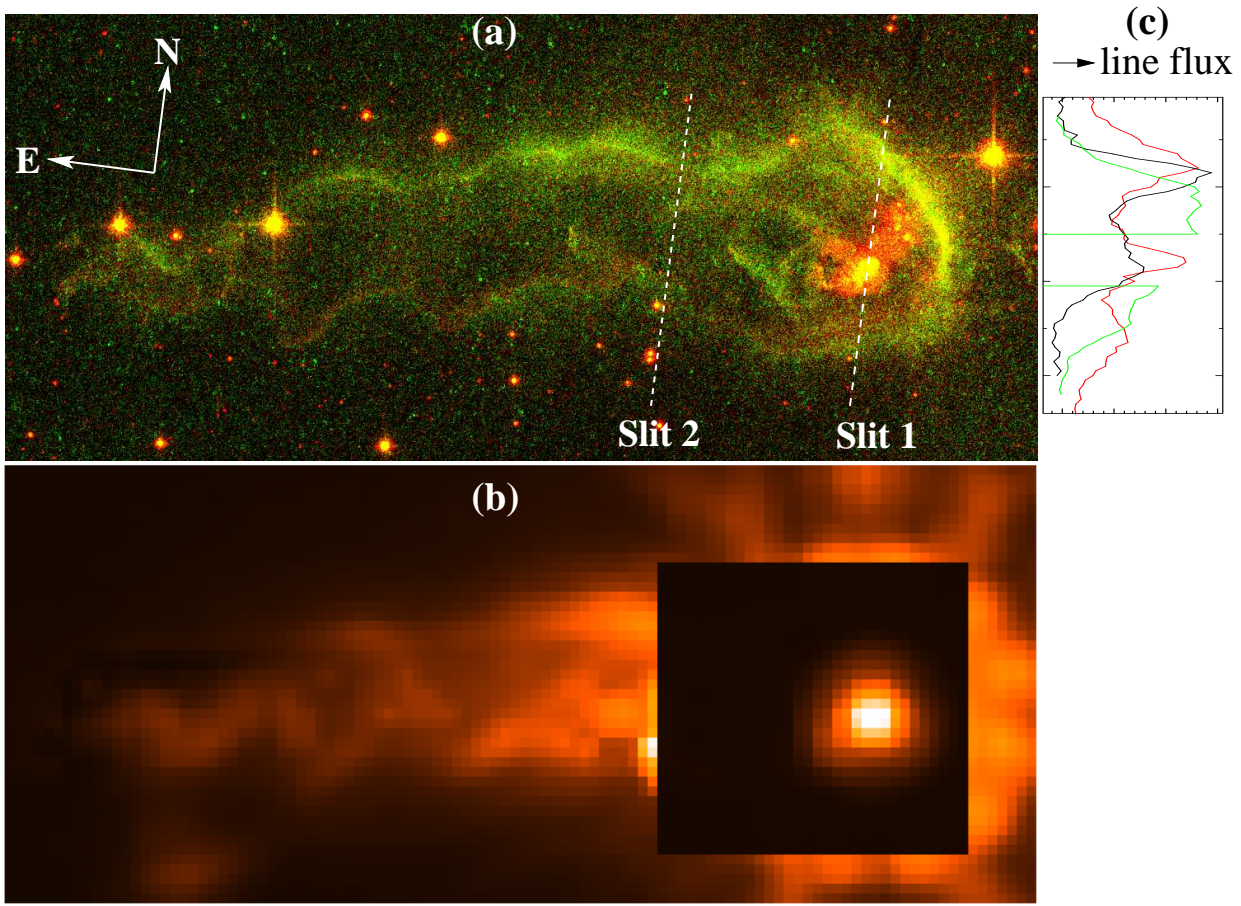}}
\caption{(a) Composite HST ACS/WFC (false-color) image (size $62.7\arcsec\times27.1\arcsec$) of I\,20324 taken through two
broad-band filters F606W ({\it green}) and F814W ({\it red}). (b) False-color IRAC 8\,\micron~image of same field-of-view as in panel $a$:   
the intensity in a $19.2\arcsec\times18\arcsec$ box centered on the central star of I\,20324 has been scaled by a factor 0.01 in 
order to clearly show the much fainter tail structure.  Linear stretches have been used for the images in these panels. (c) Spatial intensity cuts of near-infrared emission lines as seen through a
$1\arcsec$ wide slit, at two locations towards I\,20324 -- Slit\,1:\,({\it green})\,$2.1\,\micron$ H$_2$\,v=1-0,\,S(1) 
\& ({\it red})\,$1.28\,\micron~$Pa\,$\beta$; Slit\,2: ({\it black})\,H$_2$\,v=1-0,\,S(1). The continuum emission from the central source has
been subtracted in the
Slit\,1 cuts: the intensity of  the H$_2$ line in the region of the central source (which is very bright at $2\,\micron$) has been
masked out because the residuals from the subtraction process are much larger than the line intensity.
}
\label{i20324hst_h2}
\end{figure}

\begin{figure}[!h]
\resizebox{1.0\textwidth}{!}{\includegraphics{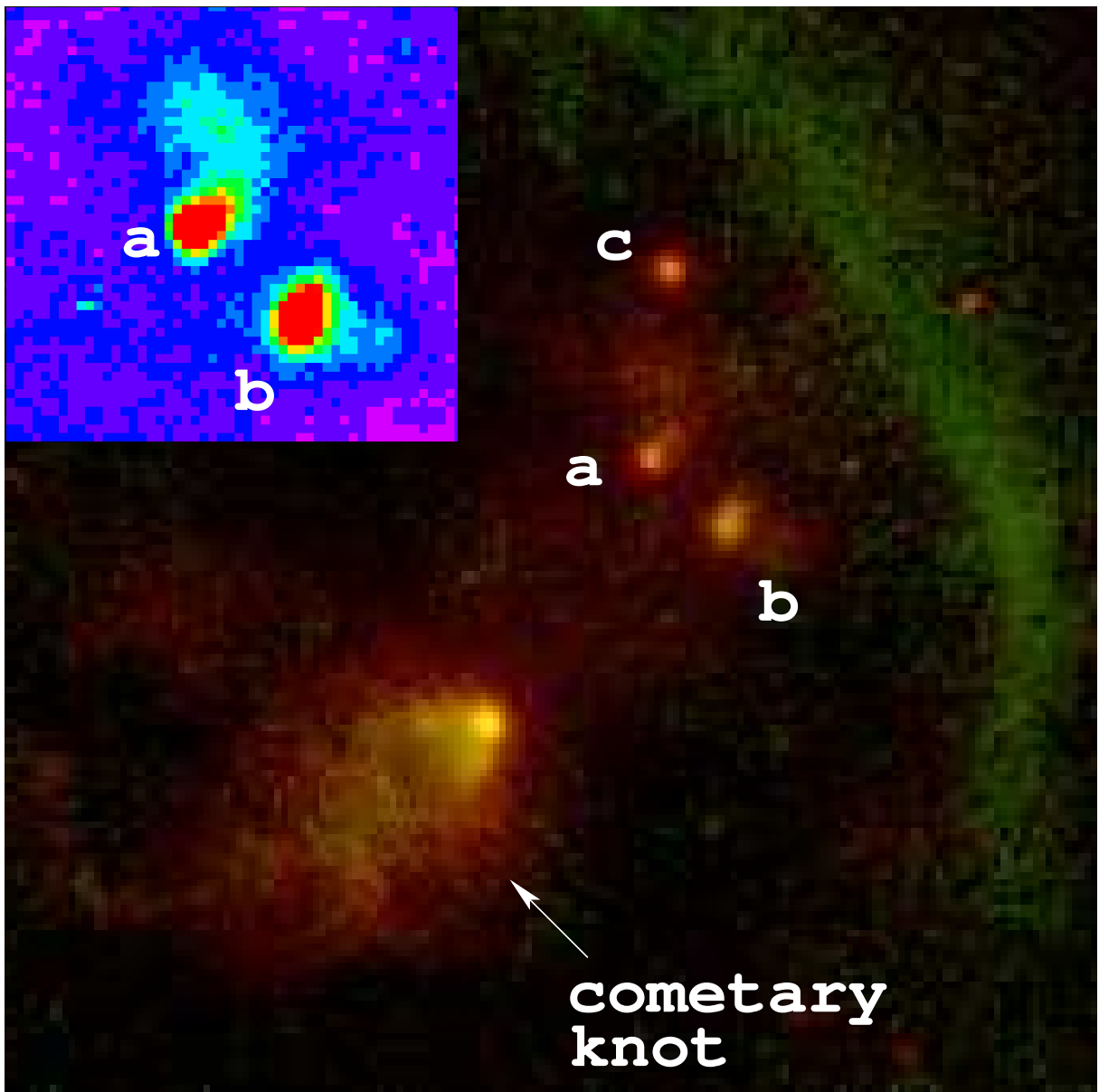}}
\caption{As in Fig. 1(a), but for a $8.5\arcsec\times8.5\arcsec$ region covering the cometary knot (a logarithmic stretch has been used to display 
each of the F606W and F814W images in the color composite). Inset shows the F814W image of the region around stars $a$ and $b$ in false-color to emphasize the faint tail structures.
}
\label{i20324hstcen}
\end{figure}

We obtained optical images of I\,20324 on 2006-07-22 using HST's Wide-Field Camera (WFC)
of the Advanced Camera for Surveys (ACS), which has a plate scale of 0\farcs050/pixel, using 
the F606W ($\lambda=0.60\,\micron$, $\Delta\lambda=0.123\,\micron$) and F814W ($\lambda=0.80\,\micron$,
$\Delta\lambda=0.149\,\micron$) filters. A 2-point dither was used for the imaging with a total exposure time of 
$2\times347$\,s per filter. These images show an extended nebula  
with a limb-brightened ``tadpole" shape (hereafter, we often use the label ``Tadpole" 
to refer to I\,20324)
% give total extent in arcsec of tail, and N-S extent of Wind Shock) 
(Fig.\,\ref{i20324hst_h2}a) oriented roughly E-W;  the periphery of the nebula is distinctly brighter and more defined
on the N side and shows prominent ripples. A bright compact knot is located interior to the tadpole head 
(Fig.\,\ref{i20324hstcen}). The intensity of the knot, which itself has a cometary shape, peaks at J2000 $\alpha$=20:34:13.23,
$\delta$=41:08:14.6 (in the F814W image), coinciding with the location of luminous point sources in the 2MASS (20341326+4108140:
$\alpha$=20:34:13.26, $\delta$=41:08:14.07) and MSX6c catalogs (G080.1909+00.5353: $\alpha$=20:34:13.4, $\delta$=41:08:14). The
axis of the cometary knot is aligned at a PA of $-43$\arcdeg (north through east). From the F606W image, we find that the Tadpole has a length of $54.7\arcsec$ (76,600\,AU) and a maximum width of $14.1\arcsec$ (19750\,AU). The cometary knot, seen more sensitively in the F814W image, is about 
$2.4\arcsec\times1.3\arcsec$ (3350\,AU$\times$1800\,AU). The knot has a conical shape near the apex with an 
``inner" opening angle of 55\arcdeg, but then narrows towards a  
more cylindrical shape at offsets greater than about $0.75\arcsec$ implying an ``outer" opening angle of $\sim0$.

The cometary knot has a diametrically-opposed faint counterpart seen only in the F814W filter -- the simplest interpretation of
these features together is that they represent scattered light from the lobes of a collimated bipolar outflow directed along the
polar axis of a flared disk (or dense equatorial region) tilted such that the near-side of the disk lies to the NW of the bright apex of the knot. Extinction
by the tilted disk then explains the faintness of the NW lobe of the bipolar outflow. The NW outflow extends toward three faint 
red star-like objects ($a$, $b$ and $c$ in Fig.\,\ref{i20324hstcen}): high-resolution ($0.15{''}$) H and K$_s$ band images obtained with the 
Laser Guide Star Adaptive Optics system on the Palomar 200-in using the High-Angular Resolution Observer (PHARO) NIR camera show point sources at the locations of $a$, $b$ and $c$ (Sahai et al. 2012, in preparation) supporting the idea that these are stars, and not simply compact disks illuminated by starlight from the Tadpole's central star. The stars $a$ and $b$ show small ``tail" structures emanating from them (inset: Fig
\,\ref{i20324hstcen}); the inner part of star $a$'s tail is roughly aligned with the bipolar outflow axis, whereas in the outer
parts, the tail starts curving backwards.

I\,20324 falls within the region imaged by the Spitzer Cygnus X Legacy Survey program (PI: J. Hora) with the IRAC (4 bands 
at 3.6, 4.5, 5.8, and 8.0\,\micron) and MIPS instruments (2 bands at 24 and 70\,\micron) -- since some of these data have been recently 
reported by Wetal12, we do not discuss them in detail. 
Inspection of the Legacy Cygnus-X Spitzer images reveals a bright point-source in all four IRAC bands and
the MIPS 24\,\micron~band, together with a tail structure. The coordinates of the point source in the Cygnus-X catalog are
RA=20:34:13.23, Dec=41:08:14.1, virtually identical with the location of the cometary knot apex in the HST images.  The 
8\,\micron~image is notable in showing the ripples on the South periphery of the Tadpole (Fig.\,\ref{i20324hst_h2}b), which are also seen in the HST image but less clearly. Although the angular 
resolution is rather limited relative to the object's size at 70\,\micron, the MIPS
image shows a cometary shape with a bright head which is likely dominated by flux from the central source in the Tadpole. 
We assume that I\,20324 
is associated with the Cygnus OB2 association, and adopt a distance, $D=1.4$\,kpc (Rygl et al. 2012)\footnote{Previous distance estimates to Cyg OB2 
have been generally higher, with $D\sim$1.7\,kpc (Hanson 2003); derived values of masses and luminosity in this paper scale as $D^2$.}.

\subsection{Additional Multi-wavelength Observations}
Following the HST observations, we obtained supporting ground-based observations with (i) the Palomar 5-m (ii) the Very Large Array (VLA) of the 
NRAO\footnote{the National Radio Astronomy
Observatory is a facility of the National Science Foundation operated under cooperative
agreement by Associated Universities, Inc.}, and (iii) the 10-m (SMT, Mt. Graham, AZ) and 12-m (Kitt Peak, AZ) (sub)millimeter-wave telescopes
of the Arizona Radio Observatory (ARO).

The TripleSpec Spectrograph at Palomar was used to observe I\,20324 with its $1\arcsec\times30\arcsec$ slit aligned N-S, so as to
cut laterally across the Tadpole body, at two locations (i) passing through the cometary knot (Slit\,1), and (ii) $12\arcsec$ E
of the knot (Slit\,2). The seeing was about $1\arcsec$ at K-band. 
Strong emission from H$_2$ S(1), v=1--0 and other H$_2$ lines was detected in both slits; for Slit\,2,
the emission is clearly limb-brightened (Fig.\,\ref{i20324hst_h2}b). The Slit\,1 spectra also show a strong continuum at the
position of the cometary knot, and extended emission
in the HI Br$\gamma$ and Pa$\beta$ lines -- the latter lines display a local peak centered on the knot.

\begin{figure}[!h]
\rotatebox{0}{\resizebox{1.0\textwidth}{!}{\includegraphics{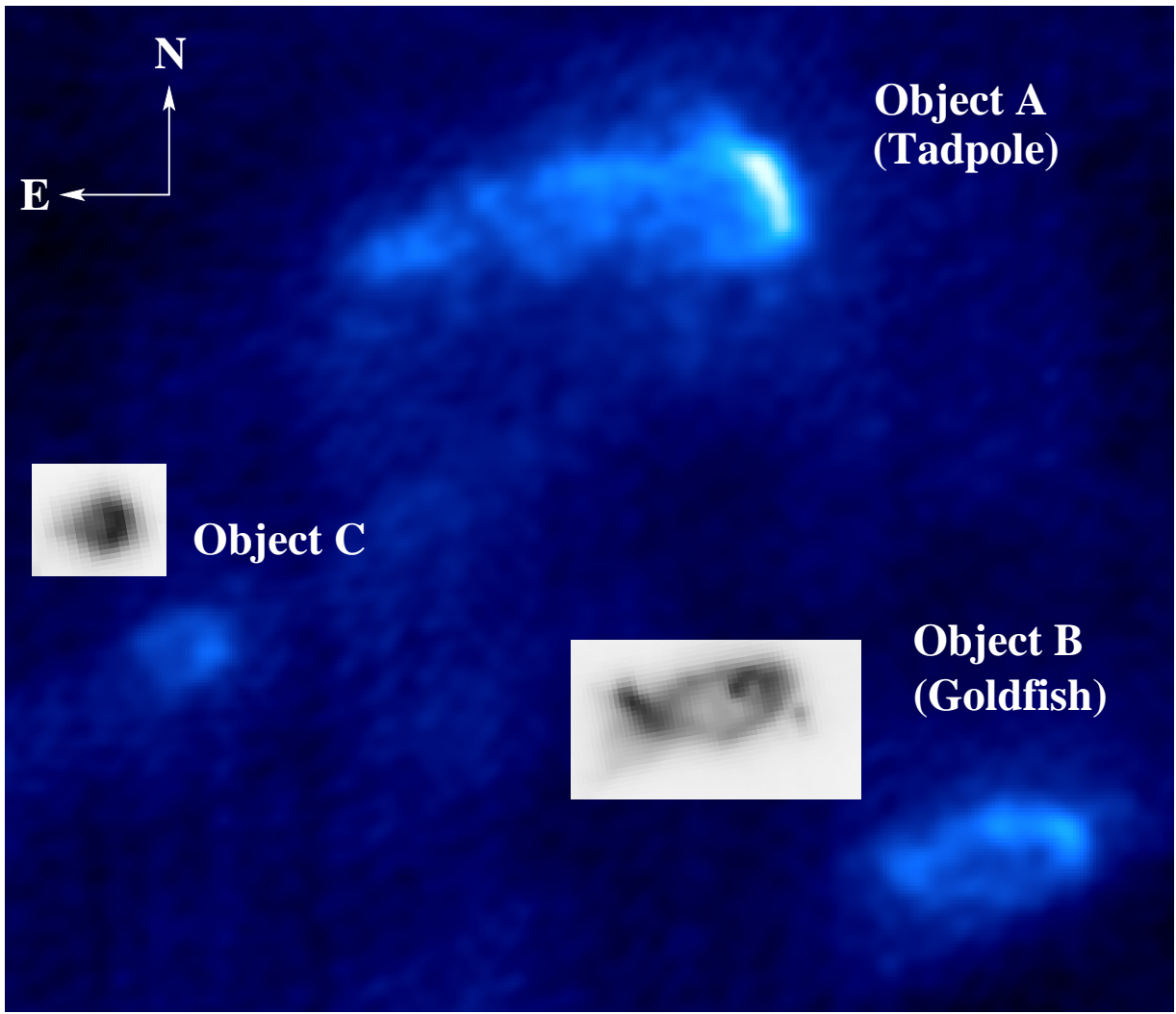}}}
\caption{VLA map of the radio emission at 8.5\,GHz from I\,20324. Panel size is $137.7{''}\times111.6{''}$. The beam is $3.2\arcsec\times2.8\arcsec$ and the rms noise 
is $\sim$65 $\mu$Jy beam$^{-1}$. The peak radio intensities towards the Tadpole, Goldfish and Object C are 1.33, 0.90, and 0.57\,mJy beam$^{-1}$. Insets show IRAC 8\,\micron~images of objects B and C, on the same angular scale.}
\label{i20324radio}
\end{figure}

The VLA was used in the {\bf C}
configuration to observe the field near the Tadpole 
nebula in August 2009, as part of program AS980.  Observations were made at frequencies 8.5 and 22.5\,GHz ($\lambda$=3.6 and 1.3
cm, respectively).  Standard data editing, calibration, imaging, and 
deconvolution were performed using the Astronomical Imaging Processing System 
(AIPS) for both frequencies.  For the 8.5\,GHz image, observed on 05\,August (Fig.\,\ref{i20324radio}), 
the restoring beam was $3.2\arcsec\times2.8\arcsec$ at $PA=-65\arcdeg$.  The rms noise 
in the final image was $\sim$65 $\mu$Jy beam$^{-1}$.  For the 22.5 GHz image (observed
on 10 August), a $70\,k\lambda$ taper was applied in the imaging step to provide a similar 
angular resolution to the 8.5\,GHz image.  The VLA 8.5\,GHz image shows a tadpole-shaped
structure at the location of I\,20324 and two additional sources, which were also detected in the PM07 emission-line survey 
(objects B \& C in their Fig.\,1; object A is I\,20324). IRAC 8\,\micron~images (insets in Fig.\,\ref{i20324radio}) of objects B \& C 
show similarly elongated morphologies: given its resemblance to a goldfish, we dub object B, the ``Goldfish". We also searched for the water maser 
emission line at 22.235\,GHz towards the Tadpole, but did not detect any emission over a total $V_{lsr}$ range of 
$\pm41.85$\,\kms~to a $1\sigma$ sensitivity of 25 mJy beam$^{-1}$ per 48.83\,kHz wide channel.

The total flux densities at 8.5\,GHz for the Tadpole, Goldfish and object C are $55\pm1.2$, $10\pm0.4$, and $6\pm0.7$\,mJy. At 22.5\,GHz, only the Tadpole is 
detected with a flux density of $30\pm1.4$\,mJy. The significantly lower flux density at 
22.5\,GHz compared to that at 8.5\,GHz for the Tadpole implies the presence of non-thermal emission in this source. The radio emission peaks strongly along the shock/ionization front at the head of the Tadpole, possibly as a result of a compressed magnetic layer in this front that is interacting with cosmic rays (CRs) associated with the Cyg OB2 association -- we note that the Fermi Large Area Telescope has recently revealed a 50-parsec wide cocoon of freshly accelerated CRs that fill the cavities carved in the Cygnus star-forming region by stellar winds and ionization fronts (Ackermann et al. 2011).  Further observations at different radio wavelengths are needed to confirm and explore the nature of non-thermal emission from the Tadpole. 

We estimate a total mass of 
ionized gas in the Tadpole of $\lesssim10^{-2}$\,\ms, assuming optically-thin free-free emission at 
22.5\,GHz, and approximating the emitting region with a triangular slab (of area 375 arcsec$^2$), which fits the Tadpole's projected shape in the 8.5\,GHz image. Our mass estimate is an upper limit 
since there may be a non-thermal contribution at 22.5\,GHz.

The ARO data were taken during 2009 Jan-May. The $^{12}$CO, $^{13}$CO\,J=2--1 lines and the HCO$^{+}$\,J=3--2 line were observed with the 1\,mm 
dual-polarization
receiver employing ALMA Band\,6 SBS mixers on the SMT 10-m, and filterbanks with 1\,MHz resolution. Typical system temperatures were 240\,K (770\,K) for the CO (HCO$^{+}$) observations. The beam size was 
$\theta_b=32$\arcsec at 1.3\,mm, and pointing accuracy was estimated to be about $\pm$4\arcsec. The $^{12}$CO and CS\,J=2--1 lines were observed with the 12-m telescope at Kitt Peak 
using dual--polarization SIS recievers. Typical system temperatures were 430\,K (240\,K) for the CO 
(CS) observations, and data were recorded with the Millimeter Autocorrelator (MAC) configured to provide 48.8\,kHz resolution. The beam size was 
$\theta_b=60$\arcsec at 2.6\,mm, and pointing accuracy was estimated to be about $\pm$6\arcsec. Observations were conducted in position-switching mode using 
an off position $15{'}$ South of I\,20324 which was tested to be free of significant emission.

\begin{figure}[!h]
\rotatebox{0}{\resizebox{1.0\textwidth}{!}{\includegraphics{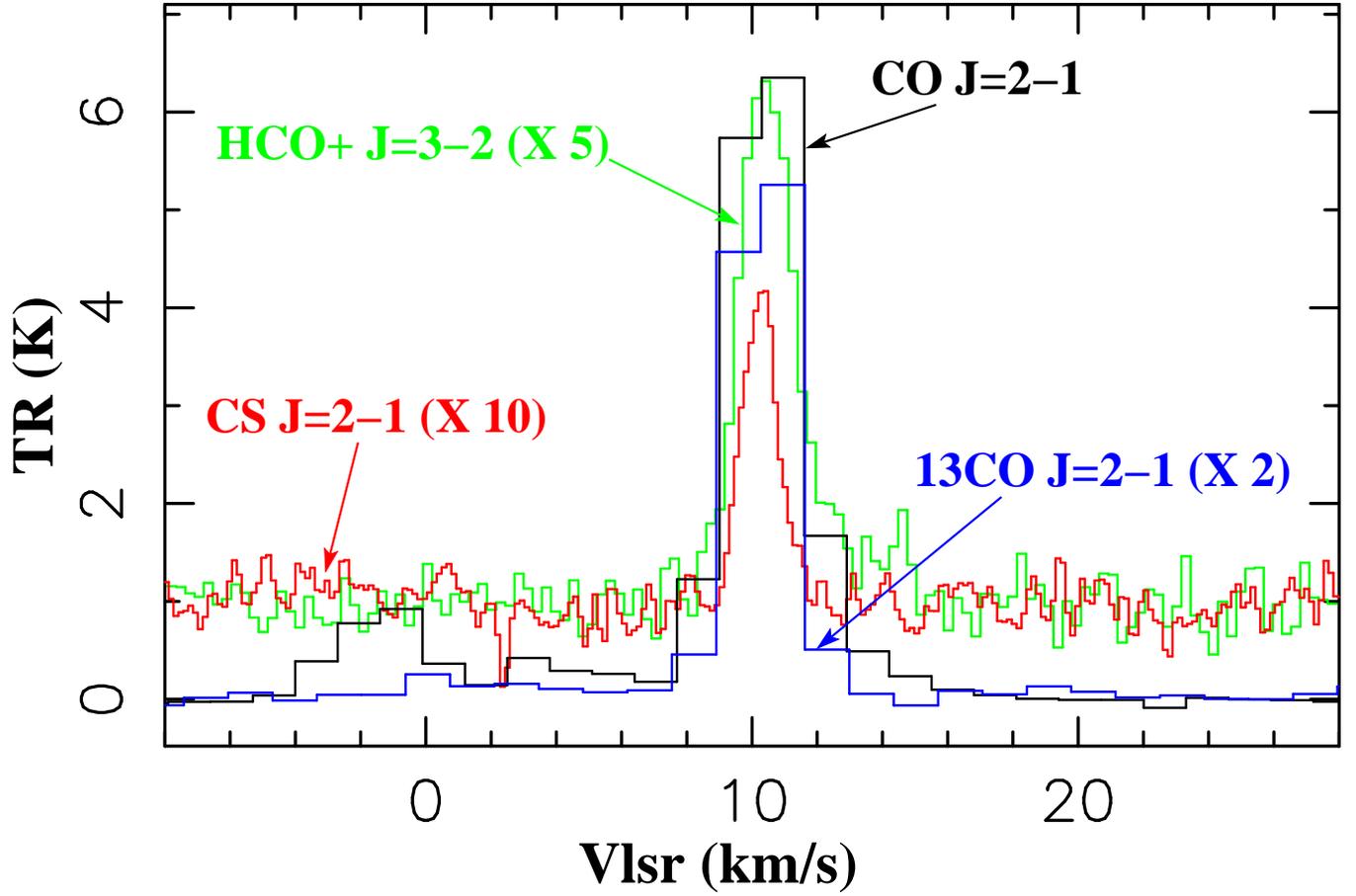}}}
\caption{Molecular line emission from I\,20324, observed with the ARO's 12-m and 10-m telescopes. For clarity, the $^{13}$CO J=2-1, CS J=2--1 and HCO$^{+}$ J=3--2 lines have been rescaled; the rescaled CS and HCO$^{+}$ lines have been shifted vertically by 1\,K. The weak CO emission centered at $V_{lsr}\sim-2$\,\kms~is due to emission from an extended molecular cloud (see Fig.\,\ref{i20324comap}).}
\label{i20324mol}
\end{figure}

\begin{figure}[!h]
\rotatebox{0}{\resizebox{1.0\textwidth}{!}{\includegraphics{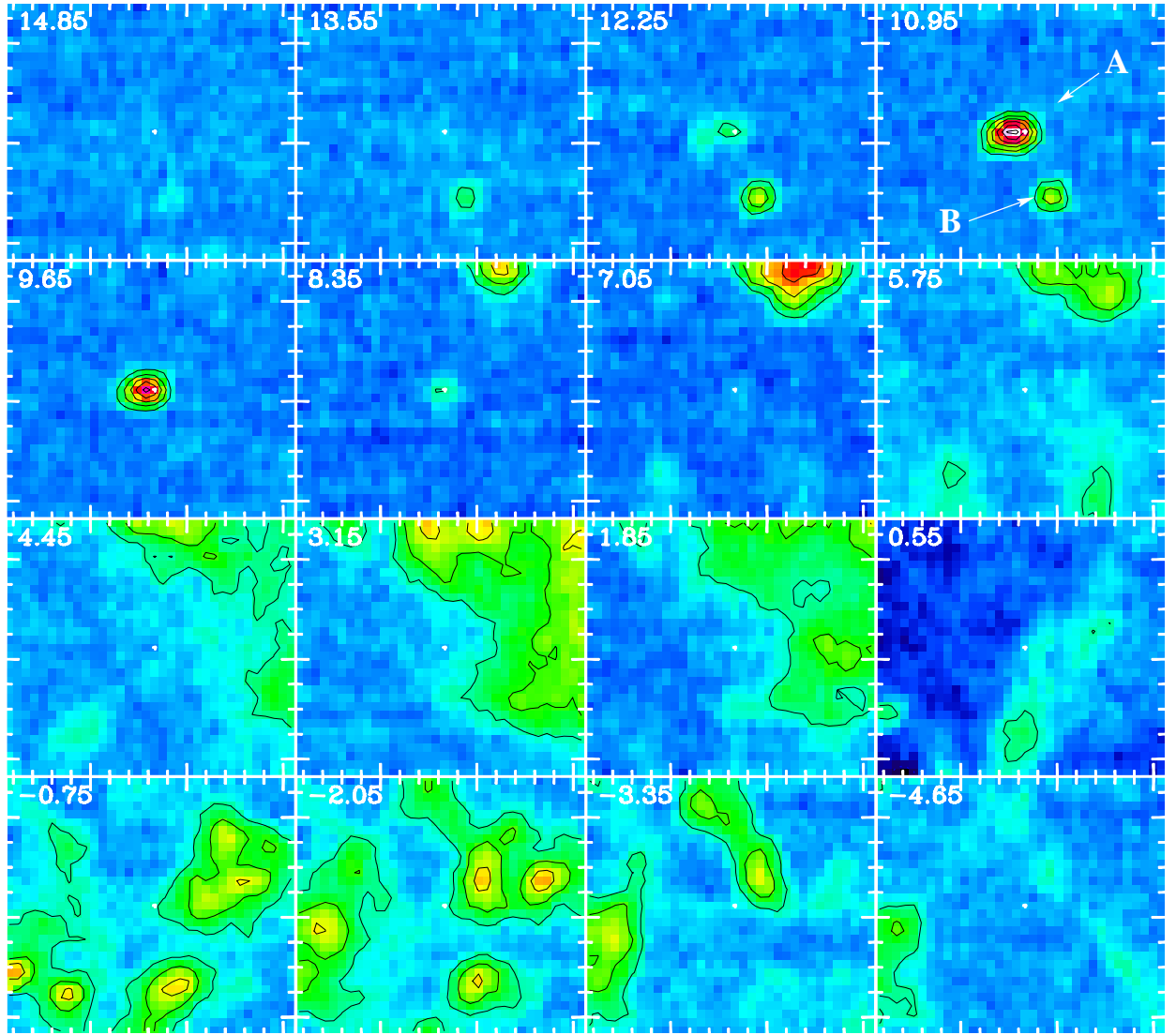}}}
\caption{Map of the CO J=2--1 emission towards I\,20324, obtained with the ARO 10-m, with each panel showing the average emission over a 1.3\,\kms~wide channel, centered at the LSR velocity shown in the top left corner. The small tickmarks on the x- and y-axes are spaced by $0.5\arcmin$ and the contours are at 1, 2, 3, 4, 5, \& 6\,K. The CO emission clumps associated with the Tadpole (``A") and the Goldfish (``B") are labelled.}
\label{i20324comap}
\end{figure}

The spectra  of the high-density tracer lines, HCO$^{+}$ J=3--2 and CS J=2--1 (Fig.\,\ref{i20324mol}), show strong emission peaking at $V_{lsr}=10.4$\,\kms, indicating that the peak density in the emitting region is quite high\footnote{The critical density for collisional excitation of the 
HCO$^{+}$ J=3--2 line is $>3.9\times10^6$\,cm$^{-3}$ in a cloud at a temperature of $<20$\,K}, $\sim\,10^6$\,cm$^{-3}$. The 
map ($5\arcmin\times5\arcmin$) of the CO and $^{13}$CO J=2-1 emission (with both lines being observed simultaneously) using the ``on-the-fly" mapping technique shows a compact source 
peaking at the same $V_{lsr}$ (Fig.\,\ref{i20324comap}) centered on the Tadpole (object A). A second nearby peak coincides with the Goldfish. 
The CO emission from I\,20324 is extended along the E-W direction with a FWHM of about
45\arcsec and it is unresolved in the N-S direction; source B is unresolved. Sparse mapping of the HCO$^{+}$ J=3--2 emission shows that it is also extended E-W.
There is also extended, structured emission in the velocity range $V_{lsr}=-5$ to $8$\,\kms~that is not connected spatially or kinematically 
with the Tadpole or Goldfish, but likely comes from the same complex of progenitor molecular clouds that spawned them.

We derive an excitation temperature, $T_{ex}=11$\,K from the peak intensity ($T_R=6.2$\,K) of the CO J=2-1 line, assuming it to be
optically thick. Assuming LTE conditions, we find (using the RATRAN online code: Van der Tak et al. 2007), that the peak $^{13}$CO J=2-1 line intensity
($T_R=2.5$\,K) and observed line-width (2\,\kms), imply a total column density $N(H_2)=0.4\times10^{22}$\,cm$^{-2}$, assuming a
standard interstellar $^{13}$CO/H$_2$ abundance ratio of $10^{-6}$ (e.g., Hayashi et al. 1993). This gives a total molecular mass of 3.7\,\ms~for I\,20324.  
Assuming a similar kinetic temperature and abundance for Source B, we find its mass is about 1.3\,\ms~(its peak $^{13}$CO J=2-1 line intensity is about half that of 
I\,20324). Our mass estimates are likely lower limits because the ``standard" value adopted for the $^{13}$CO/H$_2$ abundance ratio may be too large 
since a significant fraction of CO (and $^{13}$CO) may have been photodissociated in a PDR that occupies much of the volume of the tadpoles.

\section{The Nature of I\,20324}

\begin{figure}[!h]
\rotatebox{270}{\resizebox{0.5\textwidth}{!}{\includegraphics{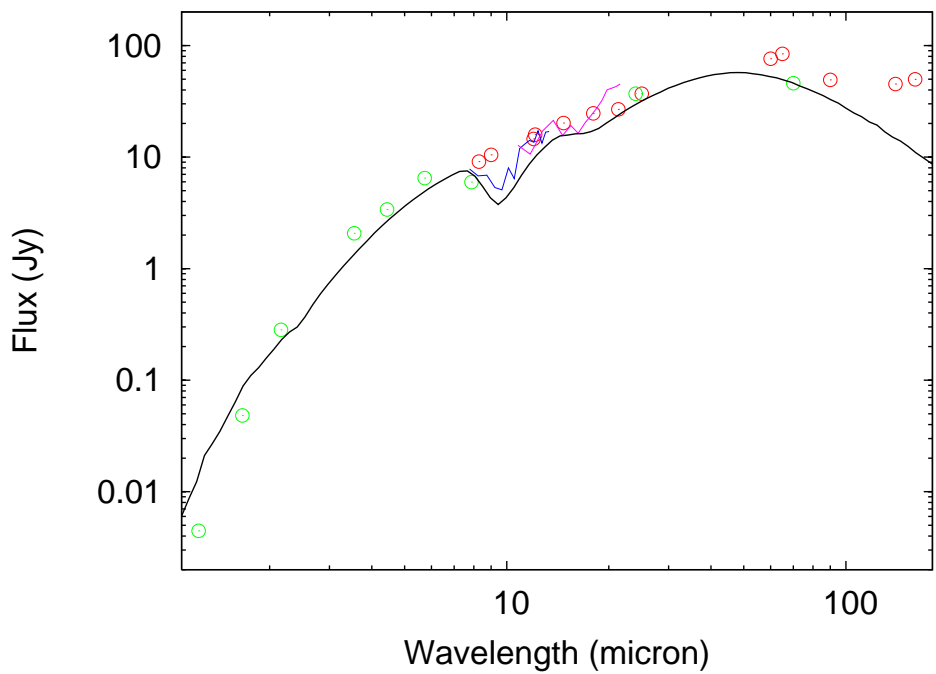}}}
\caption{The spectral energy distribution of I\,20324. The red circles show photometry from MSX, IRAS, and Akari , and the green 
circles show photometry from 2MASS and Spitzer -- the systematic errors in the photometry (not shown) are smaller than the 
symbol sizes (which correspond to $\pm15$\% errors). Blue and magenta curves show the IRAS/LRS short-wavelength 
(SW: 7.7--13.4\,\micron) and long-wavelength (LW: 11--22.6\,\micron) spectrum. The black curve is 
the SED of a pre-computed disk-envelope model for a young stellar object with a central star of mass 4.6\ms.
}
\label{i20324sed}
\end{figure}

The Tadpole is detected at all wavelengths from the near- to the far-infrared, with point-source counterparts in the 2MASS, MSX,
IRAS, and Akari catalogs: the spectral energy distribution (SED)is shown in Fig.\,\ref{i20324sed}. We have also included the IRAS Low Resolution Spectrum (LRS) spectrum for the Tadpole (with a correction of the absolute calibration applied as described by Cohen, Walker, and Witteborn (1992)). Since the LRS data were rather noisy, 
we increased the S/N ratio by binning the spectrum to a factor 3 lower resolution: a weak 10\,\micron~silicate dust feature can be seen in absorption. 
The IRAC photometry was extracted from Cygnus-X catalog, whereas the MIPS photometry was derived from the Legacy Cygnus-X images using aperture photometry. The 
core of the 24\,\micron~image of I\,20324 is partly saturated, so we used the bright Airy ring at a radius of 26.5\arcsec~from the center (4th ring) for measuring the 
flux; our estimate of 37\,Jy is consistent with the IRAS flux value. 
%from the http://www.iras.ucalgary.ca/~volk/getlrs_plot.html
Comparing the Spitzer photometry with that from missions having lower angular resolution 
({\it red} symbols) at similar wavelengths (24 and 70\,\micron), we find that the former (hereafter the ``small aperture" fluxes) fall systematically below the latter at 
$\lambda\sim60-70$\,\micron. This
indicates the presence of an extended source of warm dust at these wavelengths heated by the external radiation field, in addition to the compact central source, which looks point-like 
out to a wavelength of at least 24\,\micron. The large mid- and far-IR 
fluxes of this point-source imply the presence of a substantial mass of circumstellar dust around the Tadpole's central star. The total source luminosity of the Tadpole derived from integrating the ``small aperture" SED is about 500\ls, implying that the Tadpole's central star is not a low-mass star. From an inspection of pre-main-sequence stellar evolutionary tracks by  
Bernasconi \& Maeder (1996), we find that our estimated luminosity implies a stellar mass of around 5\ms~if the star has just begun deuterium burning, or somewhat lower if it is approaching the main-sequence.

We fitted I\,20324's ``small aperture" SED using the online tool provided by Robitaille et al.\,(2007), which computes least-squares fits of
pre-computed models of young stellar objects (YSOs) having disks and rotationally-flattened infall envelopes (with biconical outflow cavities), to user-defined SEDs. We set the input distance range to D=1.3--1.6\,kpc. The input interstellar extinction range was set to $A_v=0-2$, based on our estimate of $A_v=0.7\pm0.7$ using the numerical algorithm provided by Hakkila et al. (1997),
which computes the 3-dimensional visual interstellar extinction and its error from inputs
of Galactic longitude, latitude, and distance, from a synthesis of several published studies.
The best-fit model has 
D=1.4\,kpc, $A_v=1.5$, and a central star mass, $M_*=4.6$\,\ms~and effective temperature, $T_{eff}=4140$\,K; the model spectrum 
shows the presence of a weak silicate absorption feature as observed (Fig.\,\ref{i20324sed}). In this model, 
the full opening angle of the bipolar outflow cavity is $\theta_c=7$\arcdeg, and the outflow cavity axis is inclined at a modest 
angle of $i=18$\arcdeg, to the line-of-sight. The next best model has D=1.4\,kpc, $A_v=2$, $M_*=6.5$\,\ms, $T_{eff}=4200$\,K, 
$\theta_c=5$\arcdeg, and $i=18$\arcdeg, however the depth of the silicate feature relative to the continuum, is twice that  
observed. In both models, almost all of the emitting mass is associated with the envelope. These models are 
qualitatively consistent with the optical appearance of the luminous central source, with scattered light from the inner regions of 
the near-side bipolar cavity producing the cometary knot seen in the HST image, and its farside counterpart being 
obscured by the dense inner equatorial region of the envelope. The value of the de-projected opening angle in the best-fit model, 
$i_{corr}=tan^{-1}(tan\,\theta_c/sin\,i)=23$\arcdeg, lies in the range covered by the inner and outer opening angles of the knot ($0-55\arcdeg$). We note that the minimum inclination angle in the model is $i=18$\arcdeg, so it is possible that a better fit might have been obtained with a smaller inclination 
angle which would result in a larger value of $i_{corr}$. However, a better 10\,\micron~spectrum  covering the silicate feature (i.e., with higher S/N and a small aperture) is needed before the derived parameters from the above modelling can be put on a firm footing. 

%Envelope Cavity Angle (degrees) 3.5
% Hakkila extinction   l(deg), b(deg), d(kpc)=80.19,00.532,1.4 
%  Vis ext, err, correction, FinalValue(mag)=0.469111145  0.730981708  0.264930367  0.734041512

We note that there are several tadpole-shaped objects in the 
vicinity of I\,20324 that share with the latter a common E-W head-tail orientation. These are:
(i) the Goldfish (outside the field-of-view in our HST images; \#8 in Wetal12), 
(ii) ``object C" of PM07 seen faintly in the 8.5\,GHz, HST and Spitzer images,
which has a relatively small tail (\#9 in Wetal12), (iii) and an object
at RA=20:34:45.9, Dec=41:14:46.5 in the IRAC images of this region (\#10 in Wetal12). The common E-W orientation strongly suggests these structures are shaped by the ram pressure of a passing wind from a distant
source or sources located to the West of these objects, with material being
ablated from their heads and blown eastwards. 	
Noting that the tails in both the Tadpole and Goldfish are bright on their northern sides (both in the optical and in the radio), we conclude
that this is due to asymmetric irradiation of these objects by the ultraviolet radiation field of the Cyg OB2 No. 8 cluster that lies to 
NW of the Tadpole and Goldfish, at a separation of about 900\arcsec ($\sim$6\,pc). 

The Tadpole and its associates are most likely 
small, dense molecular cores near the Cyg OB2 complex that are being subjected to photoevaporation by the strong UV radiation field due to the
large number of O stars in this massive star formation region. The presence of
additional young stars (objects $a$, $b$, and $c$: Sahai et al. 2012, in preparation) in close physical proximity to I\,20324's cometary knot indicates that several young stars have formed within the dense
molecular core inside the head of the Tadpole. The small tails associated with $a$ and $b$ may represent (i) winds from these stars interacting with 
the NW outflow of, or (ii) possibly photoevaporation of their circumstellar disks by radiation from, the Tadpole's central star. The curvature of $a$'s  tail away from the Tadpole's head may be due to interaction with the compressed ``wall" of gas in the latter (note that the wall curves around both the back and the front, so $a$ could be located quite close to the rear Tadpole wall).  

The physical processes responsible for producing the radio-bright peripheries of the Tadpole and its associates are likely the same ones   
that operate in the vicinity of the M16 (Eagle Nebula) elephant trunks (lateral width
$\sim(0.5-1)\times10^{18}$\,cm) where dense, dusty molecular clouds are being eroded by photoevaporation due to radiation from O stars at a
distance of $6\times10^{18}$\,cm from the trunks, producing a photoevaporative flow. The heads of these trunks appear in
emission-line images as limb-brightened arcs at the interface between the molecular cloud and the ambient HII
region. The bright heads (size $\sim3\times10^{17}$\,cm) and bodies of the Tadpole and its associates likely represent similar
interfaces.

We interpret the ripples evident in the Tadpole body as resulting from Kelvin-Helmholtz (K-H) instabilities. 
Such instabilities are expected to occur at the periphery of a dense cloud embedded in diffuse gas, when there is
relative motion between the two components (e.g., Fleck 1984, Kamaya 1998). The spatial wavelength of the ripples 
($\sim15\arcsec=3.2\times 10^{17}$\,cm) 
is very similar to the wavelengths of the ``ripples" found on the surfaces of some molecular clouds in the Orion nebula by Bern{\'e} et al. (2010) that  
have also been interpreted as resulting from K-H instabilities. We consider (but discard) the possibility that the ripples could be caused by the orbital 
motion of two bound stars separated by a distance comparable to the ripple amplitude ($\gtrsim$2\arcsec=2800\,AU), since the ripple wavelength implies 
a time scale of $1050$\,yr\,($100\,\kms/V_w$) ($V_w$ is an assumed speed for the exterior wind impacting the Tadpole), which is much shorter than 
the expected orbital time for such a binary system, 
$10^5$\,yr\,$(M_1+M_2)^{-1/2}$ (taking the masses of the stars in the binary to be $M_1\sim\,M_2\sim\,1$\,\ms).

Although the bow-shock morphology of the head of the tadpole structure and the cometary shape of the central knot 
in I\,20324 are very similar to the shapes of structures seen towards 
proplyds in the Orion star-forming region (e.g., O{'}Dell \& Wong 
1996, Bally et al. 1998), we do not think I\,20234 or object B are proplyds because the associated molecular masses (greater than 3.7 and 1.3\,\ms) are much too large  
compared to what one might expect for
circumstellar material being evaporated from protoplanetary disks: disk masses of (0.003--0.07)\,\ms~have been inferred from an SMA continuum survey for the 
Orion proplyds (Mann \& Williams 2010). Indeed, no dense molecular medium has yet been found around the central disks in any known proplyd.  Furthermore, 
the minimum separation between the bright periphery of the Tadpole head (which would represent the wind shock in the proplyd hypothesis) and the cometary knot in I\,20324 is
$4.7\arcsec=6600$\,AU, is much larger than the corresponding typical separations for the Orion proplyds ($\sim500-1000$\,AU: Bally et al. 1998). However, this argument 
needs to be strengthened by a quantitative assessment of the competing effects of (i) the larger number of ionizing stars in 
Cyg OB2 and, (ii) their larger distance from the location of I\,20324 and its associates compared to Orion, since  
the physical scales of structures associated with proplyds scale inversely with the ionizing EUV flux. 

We conclude that I\,20324 and its associates are EGGs.  EGGs
are likely to be the surviving dense portions of their parent molecular clouds, and as such, they should be 
predominantly molecular, as observed in our objects. While the original concept of EGGs shows them as having elongated tails that connect them continuously with the elephant trunks in M16, this is due to the ionizing sources in M16 being concentrated in one 
direction relative to the EGGs. However, the presence of multiple ionizing sources distributed over a large solid angle relative to the EGGs, as in Cyg OB2, results in a diffuse radiation field that eventually pinches off their tails -- a possibility noted by Hester et al. (1996), and supported by numerical simulations (e.g., Ercolano \& Gritschneder 2011) -- creating ``free-floating" EGGs, such as the Tadpole and its associates. 

Their relatively high internal density has not only prolonged their 
survival in the hostile radiative and windy environment of the Cyg OB2 cluster, but it makes these globules a likely 
place for stars to have begun forming, especially as the overpressure of the surrounding HII region would have
aided gravity by compressing the globules.  It therefore seems to be no accident that the Tadpole and the 
Goldfish have embedded stars.  Wetal12 point out that a higher fraction of the globules in this
region (70\,\%) appear to have stars than those in the Eagle Nebula (15\,\%), and they use this fact to argue 
that their objects in Cygnus are therefore unlikely to be EGGs.  However, several variables can affect
this fraction.  The parent cloud of the Eagle Nebula may have been considerably less dense and massive
than the parent cloud of the globules in Cygnus, and the Cygnus OB2 association could
provide a much stronger radiation field and more powerful winds than the cluster powering the Eagle
Nebula.  Both of these factors would affect both the size distribution 
of the resulting EGGs and the compression to which they are subjected. 
%(see, e.g., Wright et al. 2010) 

In summary, it is quite plausible that not only 
the Tadpole and the Goldfish, but all of the ``proplyd-like" objects listed in Wetal12 are EGGs, 
considering their rather similar physical sizes and shapes. Molecular-line observations of these objects, like the ones reported in this paper, will 
help in determining their masses and thus confirming their true nature.

\section{Acknowledgments} We thank Anna Rosen for her valuable help in reducing the ARO on-the-fly data presented in this paper, as part of her 
Spring 2009 NASA/USRP student internship at JPL. We thank the staff of the 
Arizona Radio Observatory for granting us observing time. The National Radio Astronomy Observatory is a facility of the National
Science Foundation operated under cooperative agreement by Associated Universities, Inc. RS's contribution to the
research described here was carried out at the Jet Propulsion Laboratory, California Institute of Technology, under a
contract with NASA. Financial support was provided by NASA through a Long Term Space Astrophysics 
award to RS and MM, and HST GO award 10536 to RS.

%/u4/usr41/sahai/prog/i20324ab.gnu

\end{document}